# Simultaneous Measurements of Microwave Photoresistance and Cyclotron Reflection in the Multi-Photon Regime

Jie Zhang[1*], Rui-Rui Du[1,2], L. N. Pfeiffer[3], and K. W. West[3]

[1]*Department of Physics and Astronomy, Rice University, Houston, Texas 77251, USA*
[2]*International Center for Quantum Materials, Peking University, Beijing 100871, China*
[3]*Department of Electrical Engineering, Princeton University, Princeton, New Jersey 08544, USA*

## Abstract

We simultaneously measure photoresistance with electrical transport and plasmon-cyclotron resonance (PCR) using microwave reflection spectroscopy in high mobility GaAs/AlGaAs quantum wells under a perpendicular magnetic field. Multi-photon transitions are revealed as sharp peaks in the resistance and the cyclotron reflection on samples with various carrier densities. Our main finding is that plasmon coupling is relevant in the cyclotron reflection spectrum but has not been observed in the electrical conductivity signal. We discuss possible mechanisms relevant to reflection or dc conductivity signal to explain this discrepancy. We further confirm a trend that higher order multi-photon features can be observed using higher carrier density samples.

PACS#

78.67.-n

79.20.Ws

* jz38@rice.edu

I. Introduction

Semiconductor interband/intraband transitions are traditionally studied via optical absorption and transmission. Nonlinear processes involving multiple photons attract tremendous attention not only for elucidating band structures beyond the experimentally attainable frequency range, but also for understanding the kinetic properties of solids in the presence of strong radiation.

With millimeter wave irradiation, non-equilibrium phenomenon such as microwave-induced resistance oscillation (MIRO) [1,2] and zero-resistance states (ZRS) [3,4] were observed in two dimensional electron gases (2DEGs) over a decade ago. Theories have been advanced to explain their origin, the most popular being the displacement model proposed by [5] and the inelastic model by [6].

The demonstration of multi-photon cyclotron resonance (CR) corresponding to the rational values of $j = \omega/\omega_c = p/q = 1/2, 1/3, 2/3 ...$ (where $\omega$ is the microwave frequency, $\omega_c$ is the cyclotron frequency, p and q are integers) via microwave reflection spectroscopy [7] provides a unique method to probe this nonlinear process. Multi-photon CR is equivalent to sub-harmonic transitions where the energy of q photons matches the spacing of p Landau levels (LLs) (see Fig.2 d). Experiments have been performed to study this phenomenon, but only using transport measurements. Dorozhkin et al. [7] proposed that sub-harmonic MIROs were suppressed below 30GHz due to single photon inelastic mechanisms when LLs start to overlap, yet the experiments of Zudov et al. [8] found no frequency threshold. Wiedmann et al. observed high order MIRO at temperatures up to 6.5K [9]. Although multi-photon processes usually happen at high intensities in the terahertz and infrared regime, they also occur due to large DC fields [10]. We access this process in the absence of high AC or DC fields in a much lower frequency range where multi-photon transitions are prominent in high-mobility 2DEGs.

In present work, a simultaneous measurement of PCR reflection and MIRO are undertaken on high mobility samples with different carrier densities. The prediction of multi-photon transitions facilitated by higher density [11] is confirmed by both optical and electrical transport measurement. Power, frequency and temperature dependences are analyzed on a specific sample. The visibility of sharp signals from devices fabricated at varying widths enables us to tune the bulk plasmon frequency—a potential tool for coupling other fundamental excitations like composite fermions in the fractional quantum Hall states to photons in a detectable frequency range.

## II. Experiment

The experiments are performed in GaAs/Al$_{1-x}$Ga$_x$As quantum wells grown by molecular beam epitaxy. The specimens have carrier densities ranging from $0.8$ to $4 \times 10^{11}/\text{cm}^2$ and mobility about $10^7 \text{ cm}^2/(\text{V}\cdot\text{s})$ at 300 mK following 30 minutes of illumination under red light emitting diode. The sample is inserted into a microwave reflection interferometer (Fig 1a.). A similar schematic description of the microwave interferometer was reported in [7] where the TE$_{01}$ mode microwave in the range of 26-40GHz is transmitted and reflected through a WG28 waveguide. The incident microwave field is divided by a magic tee which isolates the two collinear ports. The sample is mounted on a thin copper plate to increase reflectivity. An adjustable short is used to tune this interference sinusoidally at a fixed frequency. A sensor is mounted to the output port for power detection, the voltage of which is then fed into a lock-in amplifier. We attach 8 indium contacts for standard transport measurement by annealing at 420℃ in forming gas. The amplitude modulation of the microwave is provided by the first lock-in amplifier at 100Hz which is much larger than 7Hz used for the second lock-in amplifier for photoresistance. A current of 1uA is applied to the sample. The transport probe is top-loaded into a He3 cryostat which operates at a base temperature of 300mK and is equipped with a 12T superconducting magnetic coil. Thus we are able to perform simultaneous measurement of photoresistance and reflection at a given microwave frequency by sweeping the external magnetic field.

The instrumental sensitivity was demonstrated in [11] revealing a visible transition under MW power as low as 10nW on both 2DEGs and two dimensional hole gases (2DHGs). We further report the success of patterning mesoscopic structures by standard photolithography on the sample without decreasing the signal/noise ratio. This is possible because the reflection setup obviates the need for sample thinning which is required in conventional transmission spectroscopy.

### III. Results and Discussions
#### A. General Considerations

Electrical transport and optical methods are common techniques to probe cyclotron transitions in a 2D system with LL spectrum of

$$E = \hbar\omega_c(N + \frac{1}{2}) \qquad (1).$$

When an integer multiple of MW frequency matches an integer multiple of the cyclotron frequency

$$\omega_c = eB/m^* \qquad (2),$$

where $B$ the magnetic field and $m^*$ is the effective mass of carriers, a resonance near the Fermi energy $\varepsilon_F$ occurs from the highest LL $|N_F\rangle \to |N_F + 1\rangle$ transition (Fig.1 (b)) resulting in a peak in both the optical and photoresistance spectrum. By least squares fitting the MIRO peaks to this linear dispersion, the effective mass of charge carriers can be extracted. It is well known that electrons in GaAs/AlGaAs quantum wells usually have an effective mass around 0.067 $m_e$ while hole effect mass varies between 0.2 $m_e$ and 0.5 $m_e$ depending on materials parameters like well width and doping.

Due to the high mobility of samples, multi-photon transitions are clearly visible. A zero-resistance state is observed in Fig.2 (f) which is identified with sub-harmonic peaks corresponding to 2 and 3-photon absorption. The reflection signal in Fig.1 (c) also has multiple peaks related to fractional mode transitions. Transport parameters such as carrier density, mobility, and scattering time $\tau_t$ are extracted from the magnetoresistance data in Fig. 1(d) while single particle relaxation time $\tau_s$ is obtained by fitting the absorption trace with a Lorentzian. The ratio $\tau_t/\tau_s$ is greater than unity for each wafer, which indicates the dominance of small angle scattering [15].

## B. The Role of Metal Contacts

Multi-photon transitions are enhanced in strong electric fields. They can be observed under intense radiation or large DC electric fields. In our experiment, since metal contacts are used for transport measurements, the existence of a strong microwave electric field near the metal contacts is possible when the contacts are exposed to radiations.

When the large electric field of a surface plasmon at an interface interacts with photons, non-linear optical effects like second harmonics are observed [19]. In order to rule out the possibility that the multi-photon phenomenon is caused by the electric field of the surface plasmon near the contacts which is relevant in many photovoltage measurements, we prepared different sample geometry for comparison. As shown in the bottom of Fig.1 (a), a large sample whose metal contacts are positioned outside of the region of MW radiation is used in comparison with a small sample cut from the same wafer whose contacts are under the radiation. No qualitative difference in terms of the sub-harmonic order in the MIRO transitions is found except additional peaks originating from the CR of electrons in the metal are observed for small samples. We further analyze the difference of cyclotron reflection absorption signal with or without metal contacts, yet only slight modification of the peak strength is present. Therefore, we conclude that metal contacts on a sample perimeter do not play an essential role in generating high harmonics or multi-photon processes in MIRO and cyclotron absorption experiment. The phenomenon of multi-photon transitions in high mobility 2DEGs is a general result of interaction between 2D electrons from the highest LL and MW field.

## C. Plasmon Coupling

Coupling between $\omega_c$ and the bulk magnetoplasmon frequency is widely observed in CR

measurements. The bulk magnetoplasmon frequency is

$$\omega_p = \sqrt{\frac{n_s e^2}{2m^*\epsilon_{eff}}\frac{N}{W}}, \quad (3)$$

where $N \in \mathbb{N}$ is the plasmon harmonic, $n_s$ is the sheet carrier density, and $\epsilon_{eff} = (\epsilon_0 + \epsilon_1)/2$ is the effective dielectric constant, which is the average of the dielectric constant $\epsilon_1$ of GaAs and $\epsilon_0$ the dielectric constant of the vacuum) [12]. The effective resonance transition manifests itself as the plasmon-cyclotron modes

$$\omega = \sqrt{\omega_c^2 + \omega_p^2}. \quad (4)$$

Parameters like effective mass and bulk magnetoplasmon frequency can be extracted by least squares fitting the resonant absorption peaks at different frequencies with this formula (Fig. 3). Note that the deduced effective mass in our samples are slightly larger than the well-known electron effective mass of $0.067m_e$ (Table I). We attribute this discrepancy to be associated with the plasmon-coupled cyclotron mode, therefore it is not necessarily equal to the band mass. In addition, we observe the odd plasmon harmonic modes where $N = 1$, 3 and 5 in low density samples (Fig.3 (a)). However, the MIRO spectrum only shows the non-coupled cyclotron mode (Fig.2). We note that there have been no theories proposed to explain the absence of the plasmon coupled modes in MIRO.

We interpret the different phenomenon observed here in electric transport and cyclotron measurement regarding essentially the same process is due to the fact that different probing methods are used. Therefore, even with the existence of all the physical processes, only the relevant part that could be coupled to the probing device would appear in certain measurements.

The difference between an optical experiment and an electrical transport experiment under microwave radiation is that the latter involves a DC electric field in addition to the rapidly oscillating AC field. Only those scattering processes that respond to the applied DC field would be relevant in the transport measurement. Note that though the microwave has a large electric field ($\sim 10^4$V/m for 1mW power without accounting for screening) compared to the DC field ($\sim 10^{-3}$V/m), it's oscillating with a very high frequency. Therefore the Fermi sphere displacement in Drude model caused by DC field dominates over AC field which could be regarded unpolarized when it interacts with electrons. This is the same with bulk magnetoplasmon whose rapidly oscillating field does not polarize in a certain direction whereas the DC electric field does. The lack of direct correlation between the magnetoplasmon and transport scattering process might be a viable explanation of the failure to observe the PCR modes in MIRO.

### D. Higher Order Multi-Photon Features in Higher Density Samples

From the linear dependence of the cyclotron frequency on the applied magnetic field in the MIRO measurements (Fig.2), transition modes with different orders could be identified. Apart from the obvious fact that the number of visible high order harmonic peaks depends on the mobility, or more precisely, the single particle quantum lifetime, the carrier density also plays an import role. In Fig.2, we show that for wafers with carrier density lower than $1.3 \times 10^{11} \text{cm}^{-2}$, no transitions with j<1 are observed whereas for densities greater than $2.7 \times 10^{11} \text{cm}^{-2}$, two photon absorption (j=1/2) is always seen and even other fractional transitions like j=3/4 are observed in our highest density sample. Therefore, we conclude that higher carrier densities increase the observable orders of the multiphoton transitions in high mobility 2DEGs.

Surprisingly, our reflection interferometer reveals more orders of multi-photon cyclotron resonance signals than photoresistance. Limited by the fundamental mode of the waveguide, the highest PCR frequency we are able to observe is 40 GHz. So a smaller magnetoplasmon frequency would leave us a wider frequency range to work with. Compared to 2DHGs, the small effective mass in 2DEGs yields a higher plasmon frequency for millimeter size samples [11]. From equation (2), we can see that this value is determined by carrier density and the width of the sample given the same type of quantum wells. Therefore, a lower density wafer (e.g. wafer a, b) is preferred when smaller size patterns are fabricated. In fact, for wafer b, resonance peaks are observed for the entire frequency range of 26-40 GHz on the 200um wide stripe sample while when we push the plasmon frequency even higher by narrowing the width to 100um, only 3 data points close to 40GHz are collected (red dot in Fig.3 (d)). Therefore, a wafer which has lower density allows us to observe more coupling modes. For high density samples (Fig.3 (c-f)), a single magnetoplasmon frequency with different cyclotron harmonics is identified while for low density samples (Fig.3 (a)), odd plasmon harmonics (*N*=1, 3, 5) with single cyclotron mode branch out in the dispersion. The different behavior of the coupling mode still requires theoretical explanation. However, j<1 transitions are found in high-density samples using absorption, and much richer dispersion relations are apparent. Unlike transport data where the major peaks always correspond to the j=1 mode, PCR data shows a rather different manner where fractional j<1 transitions could well dominate over the fundamental cyclotron peaks.

Figure 4 shows a typical reflection signal at a fixed frequency and power on a sample cleaved from wafer e. Since the mobility is so high that peaks corresponding to different harmonic orders are well separated, Lorentzian shape fitting is possible for major features. Single particle relaxation time with each transition can be extracted from the FWHM of each peak. The wide j=1/3 peak might associate to the process of subsequent absorption of two photons after the first one before the system relaxes to the ground state, therefore relaxation time is almost 3 times that of j=1. The even smaller $\tau_s$ for j=3/4 or 4/5 indicates overlapping LLs which is reasonable at low magnetic field.

The fact that multi-photon processes and MIRO signal strength favor higher carrier density is widely observed experimentally. It can be simply understood as more electrons per unit area generate relatively higher signals. However, if we assume this nonlinearity comes from the electron-electron interaction, the ratio of Coulomb energy and kinetic energy is actually higher at low densities. Therefore, a proper theoretical framework is still needed to address the density dependence when multi-photon transitions are included in the picture.

### E. Frequency, Power, and Temperature Dependences

On wafer c where two photon transitions are significant, we perform microwave power, frequency and temperature dependence of PCR and MIRO signals. At a fixed frequency, both the signals increase with microwave power (Fig.5 (a, b)) but in a rather different manner. For MIRO, the major peak height (j=1) saturates at high microwave power which is consistent with the findings of [13] while the j=1/2 peak is hardly visible with low radiation power. However in the case of reflection absorption observed here, the dominant peak always corresponds to the two-photon absorption which has no onset power threshold and the peak heights depend linearly on the input power.

Figure 5 (c) and (d) illustrate magnetoresistance and cyclotron absorption measurements at different frequencies with a fixed microwave power. Two-photon MIRO peaks are more pronounced at lower frequencies. This is a reasonable result because the number of photons is inversely proportional to the frequency at the same power. However, this phenomenon is not observed in the reflection absorption where the major peak is always related to the two-photon transition in the frequency range. In spite of the fact that there is no frequency threshold for two-photon MIRO observed in our experiment, we cannot exclude the possibility that for separated LLs, there exists a higher frequency upper threshold of the form $\omega_c\left[1 - \arccos(1 - \alpha) - \sqrt{(2-\alpha)\alpha}\right], (\alpha < 2)$ [9].

For a fixed microwave frequency and power, increasing temperature will smear the MIRO peaks but the reflection absorption line shape remains the same within the experimental temperature range (Fig.5 (e, f)). SdHO, MIRO and cyclotron absorption are visible at different temperatures as their mechanisms are distinct. Within the investigated temperature, the CR absorption line shape remains unchanged as long as the temperature is low enough for electrons to undergo circular motion ($\omega_c \tau_q \gg 1$). SdHO can only be resolved at a lower temperature and a higher field when the LL separation ($\hbar\omega_c$) is larger than smearing of the Fermi surface ($k_B T$). It has a form

$$\delta\rho = 4\rho_0 X_T \lambda \cos(2\pi\nu)/\sinh(X_T), \quad (5)$$

where $X_T = \frac{2\pi^2 k_B T}{\hbar\omega_c}$. In order to see MIRO, other than conditions above, strong microwave radiation is required. The contribution to the photoresistance is

$$\delta\rho_i = -4\pi\rho_0 P_\omega^0 \frac{\tau_i}{\tau_{tr}} \epsilon_{ac} \lambda^2 \sin(2\pi\epsilon_{ac}) \quad (6)$$

in the overlapping LL regime ($\omega_c \tau_q < \pi/2$), where $\tau_i = 3\tau_q^{im}$ in the displacement model and $\tau_i = \tau_{in}$ in the inelastic model. Here $\tau_q^{im}$ is the long range impurity scattering time and $\tau_{in}$ is the electron-electron inelastic scattering time contribution to the single particle quantum lifetime $\tau_q$.

### F. Discussion of Simultaneous/Step-wise Transition in MIRO

When considering multi-photon processes, it is not known whether photons are absorbed simultaneously or stepwise. Dmitriev et al. [17] proposed that for well-separated LLs, sub-harmonic processes in MIRO are dominated by multi-photon inelastic mechanisms. However, microwave-induced spectral reconstruction of density of states shows single photon transitions are possible when Landau levels start to overlap.

The underlying physics of these mechanisms are rather different. For a simultaneous absorption of m (m ≥ 2) photons, the corresponding contribution to the photoresistance is $W_\pm^m$, where

$$W_\pm := \frac{\tau_q}{\tau_{tr}} \left[\frac{e\varepsilon v_F}{\hbar\omega(\omega_c \pm \omega)}\right]^2 \quad (7)$$

(+ and − correspond to left and right circularly polarized radiation, respectively) [9]. While in the framework of stepwise transitions of m single photons, the ratio of inelastic photoresistance and DC resistance is

$$\frac{\delta\rho_{in}}{\rho_D} \propto \lambda^{2m} \quad (8),$$

where $\lambda := \exp(-\frac{\pi}{\omega_c \tau_q})$ is the Dingle factor [18].

Some parameters of the contribution above could be estimated for wafer c. Transport scattering $\tau_{tr} \approx 1.3$ns and quantum relaxation time $\tau_q \approx 0.7$ps can be extracted from SdHO profiles. Note that the ratio $\tau_q/\tau_{tr}$ is roughly 100 times smaller than the value presented in [9]. The Fermi velocity is estimated at $2 \times 10^5$m/s. Due to screening, the magnitude of the electric field that the electrons experience is difficult to estimate directly. Damping effect of SdHO could be used as a reference. Since we have a small quantum lifetime, the Dingle factor is vanishingly small ($\sim 10^{-31}$). Therefore, multi-photon inelastic mechanisms seem to be more probable for sub-harmonic processes in MIRO. In addition, the quasi-linear dependence of the two-photon peak amplitude in MIRO at larger power for high frequencies indicates the significance of microwaves on the electron distribution function [6]. This is an important feature of the inelastic mechanism, further confirming our conjecture that sub-harmonic MIRO is dominated by multi-photon inelastic processes. This is consistent with one-photon CR except it saturates at a lower microwave power [6,9].

For wafer c, we also extract the single particle relaxation time $\tau_s \approx 0.3$ns responsible for LL broadening from Lorentzian fitting the cyclotron absorption line shape. It's the counterpart of but much larger than the quantum lifetime in MIRO. Within the experimental frequency range, the Dingle factor is approximately 0.96. Therefore, a model of stepwise absorptions of single photons before the electron system is completely relaxed seems more suitably--accounting for the two photon peak in reflection absorption spectra due to the long relaxation time in the strongly overlapped LL regime. The fact that two-photon peaks are higher than one-photon peaks also supports this long quantum relaxation time argument.

## G. Multi-photon Cyclotron Absorption Probability Calculation

With the constraint of energy conservation and high radiation intensity *I*, coherent multi-photon transitions between real initial and final states are allowed with the assistance of virtual intermediate states. Based on lowest-order perturbation theory, the *n*-photon absorption coefficient $K^n \propto I^n$. When intensities become high, multi-photon absorption is comparable with single photon absorption. A nonlinearity parameter $\eta_n$ could be introduced in this case,

$$\eta_n = \frac{K^n}{K^{n-1}} \quad (9)$$

where are $K^n$ and $K^{n-1}$ photon absorption coefficients and the total absorption coefficient is

$$K = \sum_n K^n. \quad (10)$$

Unlike the lowest order perturbation theory where $K^n$ is determined by the absorption of n photons, to fully develop high-order nonlinearity, virtual transitions involve absorption of (*n+m*) photons and simultaneous emission of m photons should also be considered. These absorption and emission channels interfere and suppress each other contributing to $K^n$ substantially making the transition probability more complicated [14]:

$$W^n = \frac{nK^n I}{\hbar\omega} = \frac{m_*^{\frac{3}{2}}}{\pi\hbar^4} f(n\epsilon_0)(2n\hbar\omega)^{1/2}(M^n)^2. \quad (11)$$

The high order processes are contained in the matrix element

$$M^n \approx \frac{1}{4}\sqrt{\frac{2}{5}}\left(\frac{eE}{\omega}\right)^2 \frac{n+1}{n-1}\sum_{m=-\infty}^{+\infty} J_m(\rho_2^n) J_{n-2-2m}(\rho_1^n) \quad (12)$$

with

$$\rho_1^n = \left(\frac{8n}{3m_*\hbar\omega}\right)^{1/2}\frac{eE}{\omega}, \quad \rho_2^n = \left(\frac{eE}{\omega}\right)^2 \frac{1}{2m_*\hbar\omega}\frac{n^2}{n^2-1}. \quad (13)$$

Here $f(\epsilon)$ is the distribution function and $J_m(\rho)$ is the $m$-th order Bessel function, $m_*$ is the effective mass and $E$ is the electric field. This is under the assumption that $|M^n| \ll \hbar/\tau$. $\tau$ is the lifetime of the final state. In order to achieve $\eta_n \approx 1$, intensities of tens of GW/cm$^2$ are required which is way beyond the destructive threshold. However, in a lower frequency regime, a much lower power intensity can allow this transition. We believe it is the combination of our low frequency range and high sample mobility that makes the multi-photon processes visible in our experiments.

Increasing radiation intensity obviously is not preferred since heating might alter the dielectric properties. CR in high mobility GaAs 2DEG quantum wells is a powerful approach to study multi-photon transitions because the effective mass of the carriers has a stable value and the peaks are narrow.

## IV. Conclusion

From a simultaneous measurement of photoresistance and reflection in the gigahertz regime, we observed multi-photon processes in high mobility 2DEGs where sub-harmonic transitions are identified by their magnetic field dispersion. We found that plasmon coupling is only relevant in cyclotron reflection but not MIRO. And more multi-photon orders are revealed in higher carrier density samples, a result confirmed by both transport and optical experiments. Possible explanations are discussed. For a specific type of wafer, results of frequency, power and temperature dependences are presented. We further discuss the different theories of multi-photon transition mechanisms with respect to the observations in the low frequency range.

*Acknowledgements*   The work at Rice was supported by NSF Grant (No. DMR-1508644) and Welch Foundation Grant (No. C-1682). The work at Princeton was partially funded by the Gordon and Betty Moore Foundation as well as the National Science Foundation MRSEC Program through the Princeton Center for Complex Materials (No. DMR-0819860).

Tables

| Wafer | Carrier density ($10^{11}$/cm$^2$) | Mobility ($10^6$cm$^2$/V·s.) | ($m^*$) | $\tau_t$(ps) | $\tau_s$(ps) | $\tau_t/\tau_s$ |
|---|---|---|---|---|---|---|
| **a** | 0.8 | 14 | 0.069 | 560 | 162 | 3.5 |
| **b** | 1.3 | 13 | 0.075 | 520 | 115 | 4.5 |
| **c** | 2.7 | 33 | 0.070 | 1310 | 300 | 4.4 |
| **d** | 3.6 | 4.7 | 0.069 | 187 | 53.6 | 3.5 |
| **e** | 4.1 | 11 | 0.072 | 450 | 114 | 3.9 |

Table I    Description of electron samples **a**, **b c, d, e**. $\tau_t$ and $\tau_s$ refer to transport scattering time and single particle relaxation time.

| Harmonic order j | Peak width $\Delta B$(T) | $\tau_s$(ps) |
|---|---|---|
| 1/3 | 0.0184 | 22.3 |
| 3/4 | 0.0036 | 113.8 |
| 4/5 | 0.0040 | 102.4 |
| 1 | 0.0059 | 69.4 |

Table II    Fitting parameters of reflection absorption peaks on sample cleaved from wafer e.

# Figures:

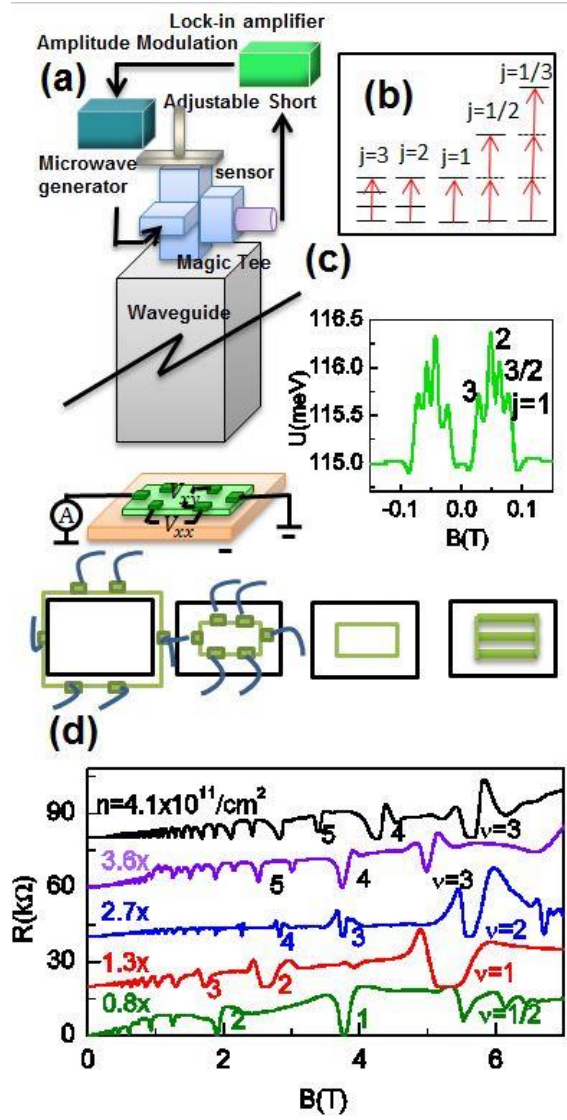

FIG. 1. (a) Experimental schematic of simultaneous measurement of photoresistance and cyclotron resonance via a reflection interferometer. (b) MIRO signal from a sample cleaved from wafer c at 31GHz and 5dbm reveal transitions containing high order harmonic and sub-harmonic processes. (c) Cyclotron reflection signal on a sample cleaved from wafer b with width of 1.7mm at 32.2GHz and 5dbm reveals multiple sub-harmonic transitions. (d) transport data shows integer

quantum hall effect (IQHE) and fractional quantum hall effect (FQHE) on sample a,b and c from which carrier density and mobility could be extracted. Symbol x denotes electron density in the unit of $10^{11}/cm^2$.

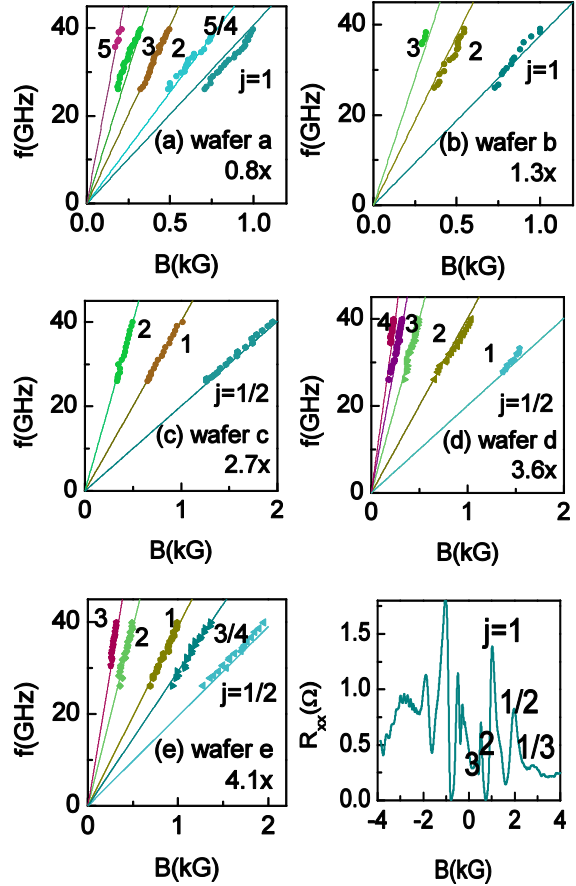

FIG. 2. (a, b, c) MIRO peaks fitting show linear dependence of cyclotron frequency with respect to the magnetic field on sample a, b and c. Insets are MIRO and SdH signals when irradiated with a fixed microwave frequency. Symbol x denotes electron density in the unit of $10^{11}/cm^2$. (d) Illustrations of high order and sub-harmonic cyclotron transitions between LLs.

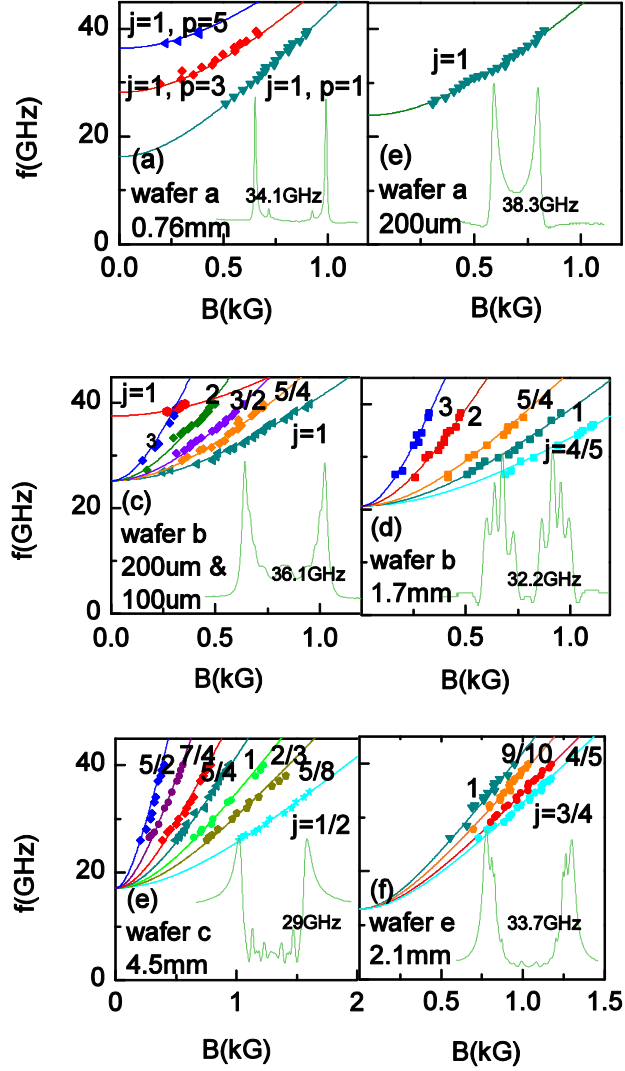

FIG. 3. (a-f) Reflection signal dispersion fitting cleaved from wafer a and patterned with different widths denoted in each figure. (g, h) Dispersion fitting on samples cleaved from wafer b and patterned with different widths.. (i) Dispersion fitting on sample cleaved from wafer c. Insets are the reflection signal line shape at a fixed frequency and 5dbm microwave power.

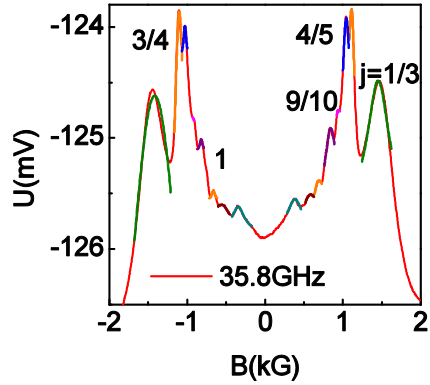

Fig. 4. A typical reflection trace (red curve) with a fixed frequency of 35.8GHz and power of 5dbm on a sample cleaved from wafer e. Peaks with different harmonic order are labeled by fractional j and fitted by a Lorentzian.

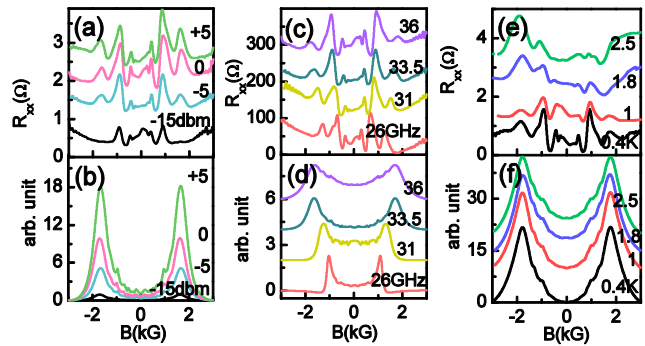

FIG. 5. (a) Power dependence of MIRO with a fixed frequency of 33.5GHz. All data traces are shifted with 0.5 for clarity. (b) Power dependence of reflection signal with a fixed frequency of 33.5GHz. All data traces are shifted by subtracting their value at -0.3T. (c) MIRO traces with selected frequencies, shifted with 100. (d) Reflection traces with selected frequencies, shifted with 2mV. (e, f) Temperature dependence of MIRO and reflection traces with a fixed frequency of 36GHz and power of 5dbm. All data traces are shifted with 1 in (e) and 10mV in (f).